\documentclass[12pt]{article}
\usepackage{epsfig}
\usepackage{relsize}
\usepackage{graphicx}
\usepackage{pdproc}
\RequirePackage{xspace}
\newcommand{\nimBaseC}       {Nucl.\ Instr.\ and Methods\xspace}
\newcommand{\jprlBase}       {Phys.\ Rev.\ Lett.\xspace}
\newcommand{\jprBase}        {Phys.\ Rev.\xspace}
\newcommand{\jplBase}        {Phys.\ Lett.\xspace}
\newcommand{\jplb}       [1]  {\jplBase\ B~{\bf #1}}
\newcommand{\nim}       [1]  {\nimBaseC~{\bf #1}}
\newcommand{\jprl}      [1]  {\jprlBase\ {\bf #1}}
\newcommand{\jprd}      [1]  {\jprBase\ D~{\bf #1}}
\def\babar{\mbox{\slshape B\kern-0.1em{\scriptsize A}\kern-0.1em B\kern-0.1em{\scriptsize A\kern-0.2em R}}}
\def\CP{\ensuremath{C\!P}\xspace}
\def\CPT{\ensuremath{C\!P\!T}\xspace}
\def\Bz{\ensuremath{B^0}}

\def\invfb{\ensuremath{\mbox{\,fb}^{-1}}\xspace}
\def\to{\ensuremath{\rightarrow}\xspace}
\def\stwob{\ensuremath{\sin\! 2 \beta   }\xspace}
\def\ctwob{\ensuremath{\cos\! 2 \beta   }\xspace}
\newcommand{\degrees}{\ensuremath{^o}}%
\def\mes        {\mbox{$m_{\rm ES}$}\xspace}
\def\Bbar    {\kern 0.18em\overline{\kern -0.18em B}{}\xspace}
\def\Bz      {\ensuremath{B^0}\xspace}
\def\Bzb     {\ensuremath{\Bbar^0}\xspace}
\def\Bu      {\ensuremath{B^+}\xspace}

\def\Bp      {\ensuremath{\Bu}\xspace}

\def\Bpm     {\ensuremath{B^\pm}\xspace}
\def\fz    {\ensuremath{f_0\!}\xspace}
\def\Kz    {\ensuremath{K^0}\xspace}
\def\KS    {\ensuremath{K^0_{\scriptscriptstyle S}}\xspace}
\def\KL    {\ensuremath{K^0_{\scriptscriptstyle L}}\xspace}
\def\Kp    {\ensuremath{K^+}\xspace}
\def\Km    {\ensuremath{K^-}\xspace}
\def\Kpm   {\ensuremath{K^\pm}\xspace}

\def\CP                {\ensuremath{C\!P}\xspace}
\def\cp                {\ensuremath{C\!P}\xspace}
\def\to                 {\ensuremath{\rightarrow}\xspace}
\def\ra                 {\ensuremath{\rightarrow}\xspace}

\def\piz   {\ensuremath{\pi^0}\xspace}

\def\BR    {\ensuremath{\Gamma}}

  \makeatletter 
  \def\@cite#1{[#1]} 
  \makeatother    
\begin{document}

\newcommand{\BABARPubYear}    {04}
\newcommand{\BABARConfNumber} {091}
\newcommand{\SLACPubNumber}{10791}
\newcommand{\LANLNumber} {0408060}

\renewcommand{\thefootnote}{\alph{footnote}}

\begin{flushright}
\babar-PROC-\BABARPubYear/\BABARConfNumber \\
SLAC-PUB-\SLACPubNumber \\
October 2004 \\
\end{flushright}

\title{
 Measurements of \CP\ Asymmetries at \babar\
}

\author{GIAMPIERO MANCINELLI}

\address{ 
Department of Physics, University of Cincinnati \\
ML 11, Cincinnati, OH 45211, USA\\
{\rm E-mail: giampi@slac.stanford.edu}\\
Representing the \babar\ Collaboration
}

\abstract{
We present preliminary measurements of \CP--violating asymmetries in $B$
decays. 
These include new results on the CKM angle $\alpha$ based on studies
of the decay $B\to \rho^+\rho^-$ and several charmonium and hadronic penguin
modes, sensitive to the CKM angle $\beta$, including results on
$B\to \phi K^0_S$, $B \to K^+K^-K^0$, $B\to \eta ' K^0_S$, $B\to f_0 K^0_S$,
and $B\to \pi^0 K^0_S$. We also report on several
of results related to the extraction of $\gamma$ and ($2\beta+\gamma$)
and present limits on \CPT violation in $B$ decays. 
}

\normalsize\baselineskip=15pt

\section{Introduction}
The unitarity of the Cabibbo-Kobayashi-Maskawa (CKM) matrix yields several 
relationships for its components, as 
${V_{ub}^\ast}{V_{ud}}+{V_{cb}^\ast}{V_{cd}}+{V_{tb}^\ast}{V_{td}}=0$.
This describes the extent of \CP\ violation in the Standard Model (SM) in
the $B$ meson system and 
can be represented in the imaginary plane as a triangle, where the
angles ($\alpha$, $\beta$ and $\gamma$) can be written in terms
of the couplings between quarks:
\begin{equation}
\alpha \equiv {\rm arg} \left [-\frac{V_{td}V^*_{tb}}{V_{ud}V^*_{ub}}\right ] ~~,~~ 
 \beta \equiv {\rm arg} \left [-\frac{V_{cd}V^*_{cb}}{V_{td}V^*_{tb}}\right ] ~~,~~
\gamma \equiv {\rm arg} \left [-\frac{V_{ud}V^*_{ub}}{V_{cd}V^*_{cb}}\right ].  \label{ckmangles}
\end{equation}
These angles can be extracted via \CP\ asymmetries 
measured in several decay modes of the $B$ meson.  We report on
recent analyses which
aim to measure these angles with data collected at the \babar\
detector~\cite{detector}. All results are preliminary unless otherwise stated.

\subsection{Measurement of $\beta$}

The angle $\beta$ can be and has been measured via time dependent asymmetry
of $B$ and $\overline B$ decays into charmonium modes. The decay rate
$\Bz\to f$, where $f$ is a \CP-eigenstate, is described by:
\begin{equation}
f_{\frac{B^0}{B^0} tag}(\Delta t) = \frac{e^{-|\Delta t|/\tau_{B^0}}}{4\tau_{B^0}}\times\left[1\mp\left(C\cos(\Delta m_{B^0}\Delta t) - 
S\sin(\Delta m_{B^0}\Delta t)\right)\right]
\label{eqn:sc}
\end{equation}
where $\Delta t$ is the time difference between the decays of  the $B$ meson studied and the other $B$ meson ($B_{tag
}$), whose decay products are used
in a partial reconstruction to infer its $\Bz$ or $\Bzb$
flavor. For charmonium modes, where the
penguin diagrams are small and carry the same weak phase as the tree
diagrams, $C=0$ and $S\propto \sin 2\beta$. \babar\ measured
$\sin 2\beta = 0.741\pm0.067\pm0.034$\cite{sin2b}. The $\sin 2\beta$
measurement assumes that the decay rates of the two mass eigenstates are
the same ($\Delta\Gamma=0$), that $q/p=1$, and that \CPT is conserved. All
these assumptions 
have been tested with a measurement which uses flavor and \CP\
eigenstates. The \CP\ sample is more important for the measurement
of $\Delta\Gamma$ as its contribution is effectively linear in this
observable, while the flavor contribution is quadratic. The results are
$|\Delta\Gamma/\Gamma|=0.008\pm0.037\pm0.018$ and
$q/p=1.029\pm0.013\pm0.011$, with all other results consistent with
\CPT conservation. Hence all measurements are 
in agreement with the assumptions made and with the SM 
predictions~\cite{cpt}. \babar\ also measures $\cos
2\beta=+2.72^{+0.50}_{-0.79}({\rm stat})\pm 0.27({\rm syst})$, 
thus a {\em positive} \ctwob value, in agreement with the SM
expectation. We estimate, using
Monte Carlo (MC), that we exclude the negative \ctwob solution at $89\%$
Confidence Level (C.L.).

The measurements of \CP\ asymmetries in modes dominated by penguin diagrams
are particularly interesting as New Physics (NP) can show up in the penguin
loops. When the tree contribution is negligible, we are effectively
measuring $\sin 2\beta$ in $b \to s$ transitions. A large departure from
the \stwob value measured with charmonium modes will indicate contribution
of NP. All the analyses on
penguin modes reported
here perform time dependent \CP\ asymmetry measurements using maximum
likelihood fits. 
In the SM, contributions beyond the leading penguin
may be difficult to estimate, depending on the channel. The ``effective
\stwob'' measured in these channels may then differ from \stwob, but
bounds on these differences are known~\cite{penthe}. 

\begin{table}
\begin{center}
\caption{\label{tab:sc}
$S$ and $C$ \CP parameters (Eq.(~\ref{eqn:sc})) measured for various $B$
  decay modes. The first uncertainty is statistical,
the second one systematic. The ``$f_{even}$'' uncertainty for $S$ of
$\Kp\Km\KS$ comes from the uncertainty on $f_{even}$ itself.} 
\begin{tabular}{|l|c|c|}\hline
$B$ decay         & $S$ & $C$ \\\hline
  $\phi\Kz$       & $+0.47\pm0.34^{+0.08}_{-0.06}$                 &
  $+0.01 \pm0.33\pm0.10$          \\
  $\Kp\Km\KS$     & $-0.56\pm 0.25\pm 0.04^{+0}_{-0.17}(f_{even})$ &
  $-0.10\pm 0.19\pm 0.09$         \\
  $\piz\KS$       & $+0.48^{+0.38}_{-0.47}\pm 0.19$                &
  $+0.40^{+0.27}_{-0.28}\pm 0.06$ \\ 
  $\fz(980)\KS$   & $-1.62^{+0.56}_{-0.51}\pm 0.09\pm 0.04(model)$ &
  $+0.27\pm 0.36\pm 0.10\pm 0.07(model)$\\ 
  $\eta'\KS$   & $+0.10\pm 0.22\pm 0.03$ &  $+0.02\pm 0.34\pm 0.03$\\\hline
\end{tabular}
\end{center}
\end{table}

\begin{figure}[htb]
\begin{minipage}{0.42\linewidth}
\begin{center}
\includegraphics*[width=1.\linewidth]{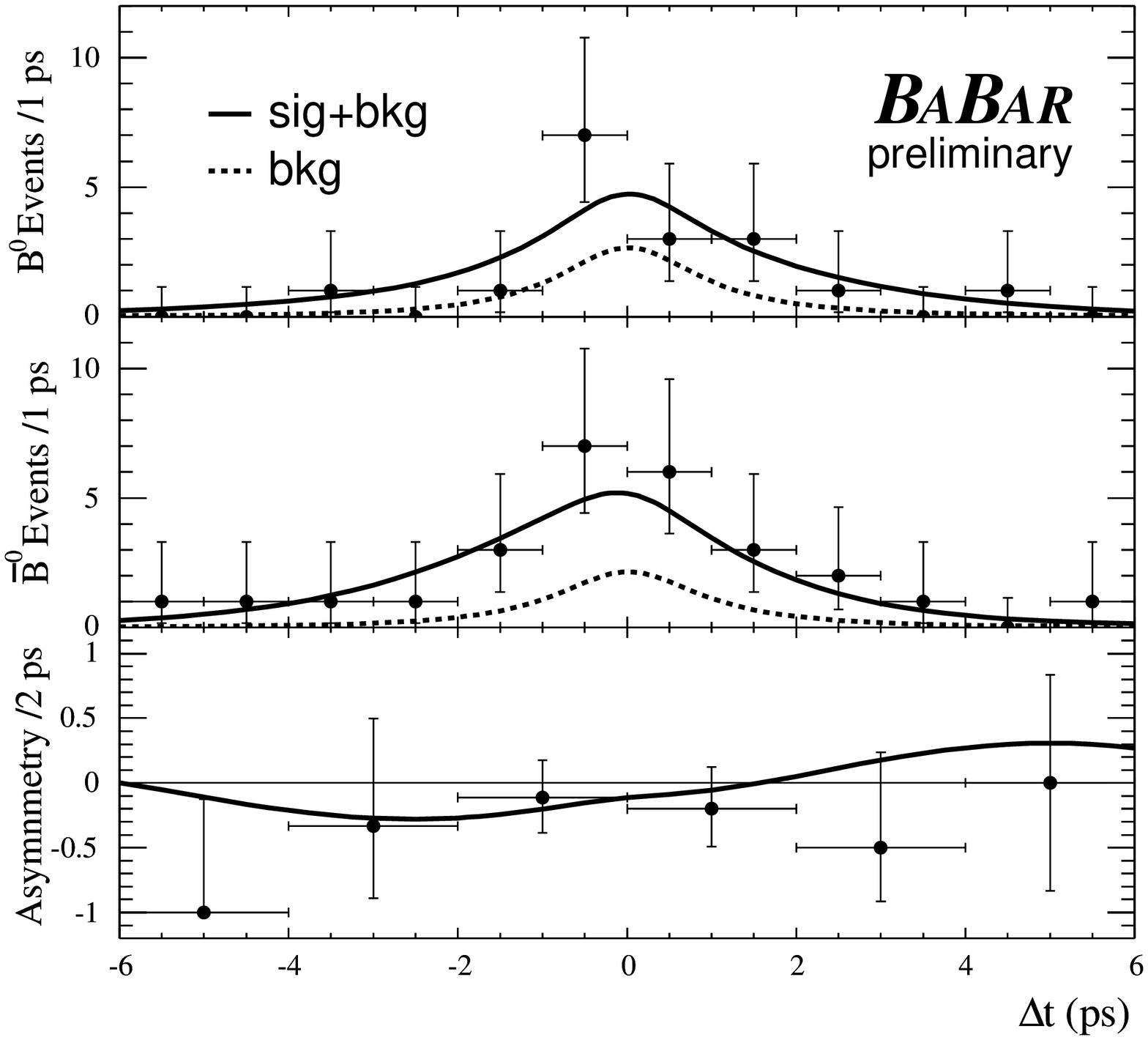}
\caption{ $\Delta t$ distributions and asymmetry of \Bz\to\piz\KS 
candidates ($122\pm 16$, found out of a 110 fb$^{-1}$ sample).}
\label{dtasymmetryproj}
\end{center}
\end{minipage}\hfill
\begin{minipage}{0.42\linewidth}
\begin{center}
\includegraphics*[width=1.\linewidth]{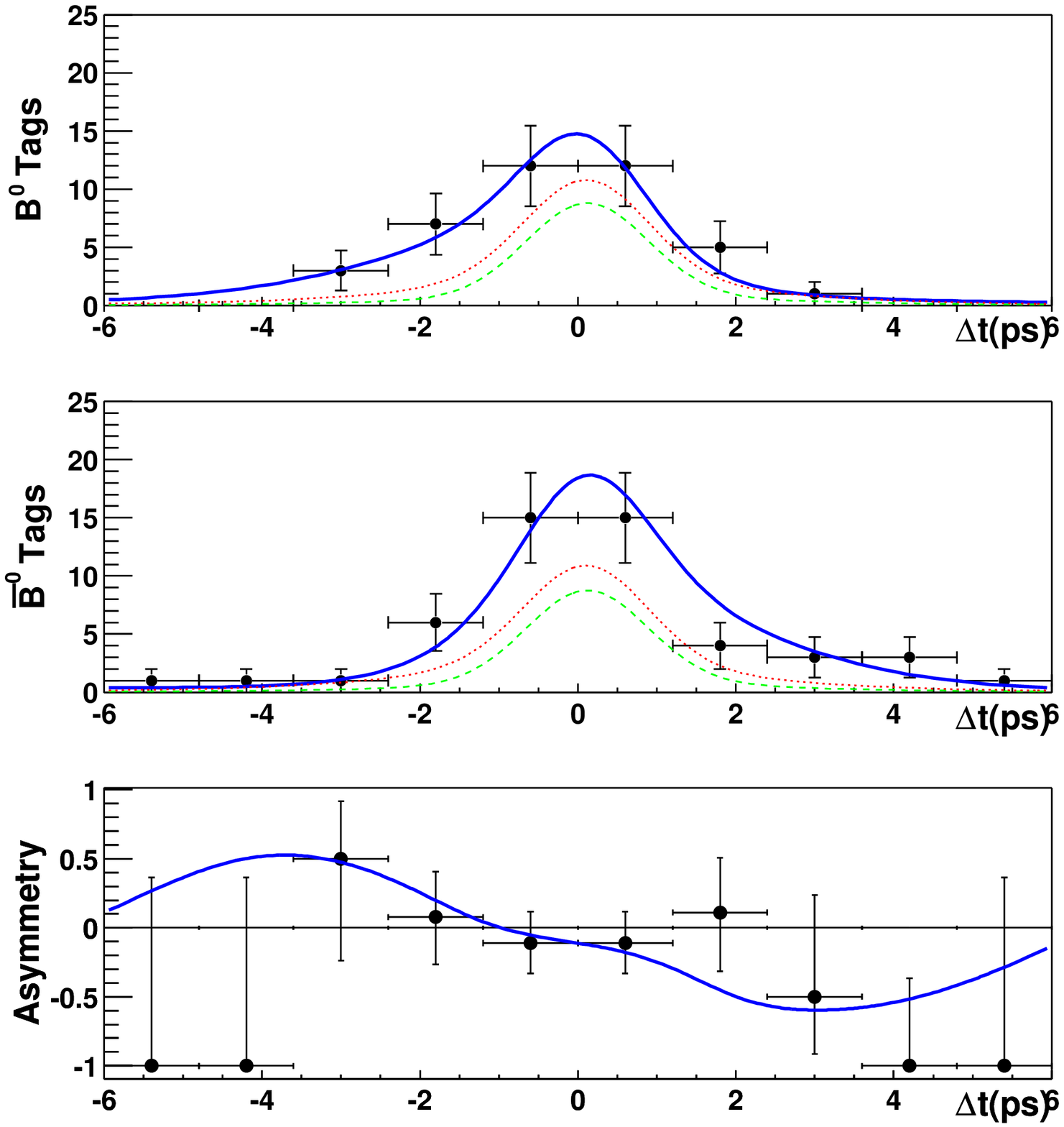}
\caption{ $\Delta t$ distributions and asymmetry of \Bz\to\fz$(980)$\KS
candidates ($94\pm 14$, found out of a 111 fb$^{-1}$ sample).}
\label{AsymmetryTot}
\end{center}
\end{minipage}\hfill
\end{figure}

The decay \Bz\to$\phi$\Kz is a $b\to s\overline{s}s$ quark level decay. In
the SM, the expected asymmetry 
$S_{\phi\KS}(S_{\phi\KL})$ is very close to $\sim +\stwob(-\stwob)$.
The \CP asymmetry parameters $S$ and $C$ measured by \babar\  
are reported in Table~\ref{tab:sc}~\cite{phiks} and are 
in agreement with the SM expectation, but \babar\ and Belle's values
present an almost 3 standard deviations (s.d.) discrepancy in the value of
$S$.
We can also measure the non resonant part of the previous decay, selecting
$K^+K^-$ pairs outside the $\phi$ mass window, and benefit from larger
statistics. 
In contrast to  \Bz\to$\phi$\KS, the \CP content is not known {\it a
priori} for this mode, but can be measured from $B\to KKK$
branching ratios (BRs) of charged and neutral $B$ mesons as:
$f_{even}={2\Gamma(\Bp\to\Kp\KS\KS)}/{\Gamma(\Bz\to\Kp\Km\KS)}$.
\babar\ measures
$f_{even}= 0.98\pm 0.15\pm 0.04$,
which is compatible with a pure \CP even state. In the SM, the expected
\Bz\to\Kp\Km\KS \CP asymmetry is then $S_{\Kp\Km\KS}\sim-\stwob$. 
The results are shown in
Table~\ref{tab:sc}~\cite{kkk}. 
\babar\ has also performed the first measurement of the \CP\ asymmetry
($A_{\CP}$) in the 
\Bpm\to\Kpm\KS\KS decay ($A_{\CP}(\Bpm\to\Kpm\KS\KS)=-0.04\pm 0.11({\rm
  stat})\pm 0.02({\rm syst})$). 
The decay \Bz\to\piz\KS has also been studied at \babar. This is a $b\to
s\overline{d}d$ quark level decay. The SM expectation for $S_{\piz\KS}$ is
$\sim+\stwob$. As we are in the 
presence of a $\pi^0$ in the final state, the position of the reconstructed
$B$ is taken constraining the $\KS$ to come from the beam spot in the
plane perpendicular to the beam direction. This requires a very good
knowledge of the beam position at all times. The results of the first
measurement of the 
\CP asymmetry for this decay are reported in
Table~\ref{tab:sc}~\cite{pi0ks}, while the decay 
rates plots are shown in Figure~\ref{dtasymmetryproj}. 
The decay \Bz\to\fz$(980)$\KS should be dominated by the $b\to
s\overline{s}s$ penguin, since the $s\overline{s} 
$ component is significant and the
$b\to u\overline{u}s$ tree is doubly Cabibbo suppressed compared to the
leading penguin. The \Bz\to\fz$(980)$\KS \CP  
asymmetry expected in the SM
is then $\sim -\stwob$.
The \CP fit result is reported in Table~\ref{tab:sc}, with decay rates
distributions shown in Figure~\ref{AsymmetryTot}. The value found for $S$
is 1.2 s.d. from the physical limit 
and 1.7 from the SM predictions.
This is the first observation of the \Bz\to\fz$(980)$\KS decay.
The results for $\Bz\to\eta'\KS$ are also reported. The $\eta'$ is
reconstructed in the $\eta\pi^+\pi^-$ and $\rho^0\gamma$ modes, with the
$\eta$ decaying into two photons and the $\rho^0$ into two charged
pions. The results are reported in Table~\ref{tab:sc}~\cite{etaprimeks}.
Combining results from all modes and from \babar\ and Belle, $\sin 2\beta$
from charmonium modes is almost 3 s.d. away from the value obtained from
penguin modes. 

\subsection{Measurement of $\alpha$}

For the main $B$ decay modes which have been investigated for the
measurement of 
$\alpha$,  $\pi^+\pi^-$ and $\rho^+\rho^-$, both tree and penguin
diagrams contribute, hence we can only measure an $\alpha$ effective. 
\babar's results with the $\pi^+\pi^-$mode are:
$C = -0.19 \pm 0.19 \pm 0.05$ and $S = -0.40 \pm 0.22 \pm 0.03$, 
which are both 2 s.d. apart from Belle's. 

\begin{figure}[htb]
\begin{minipage}[t]{8cm}
\begin{center}
\includegraphics*[width=8cm]{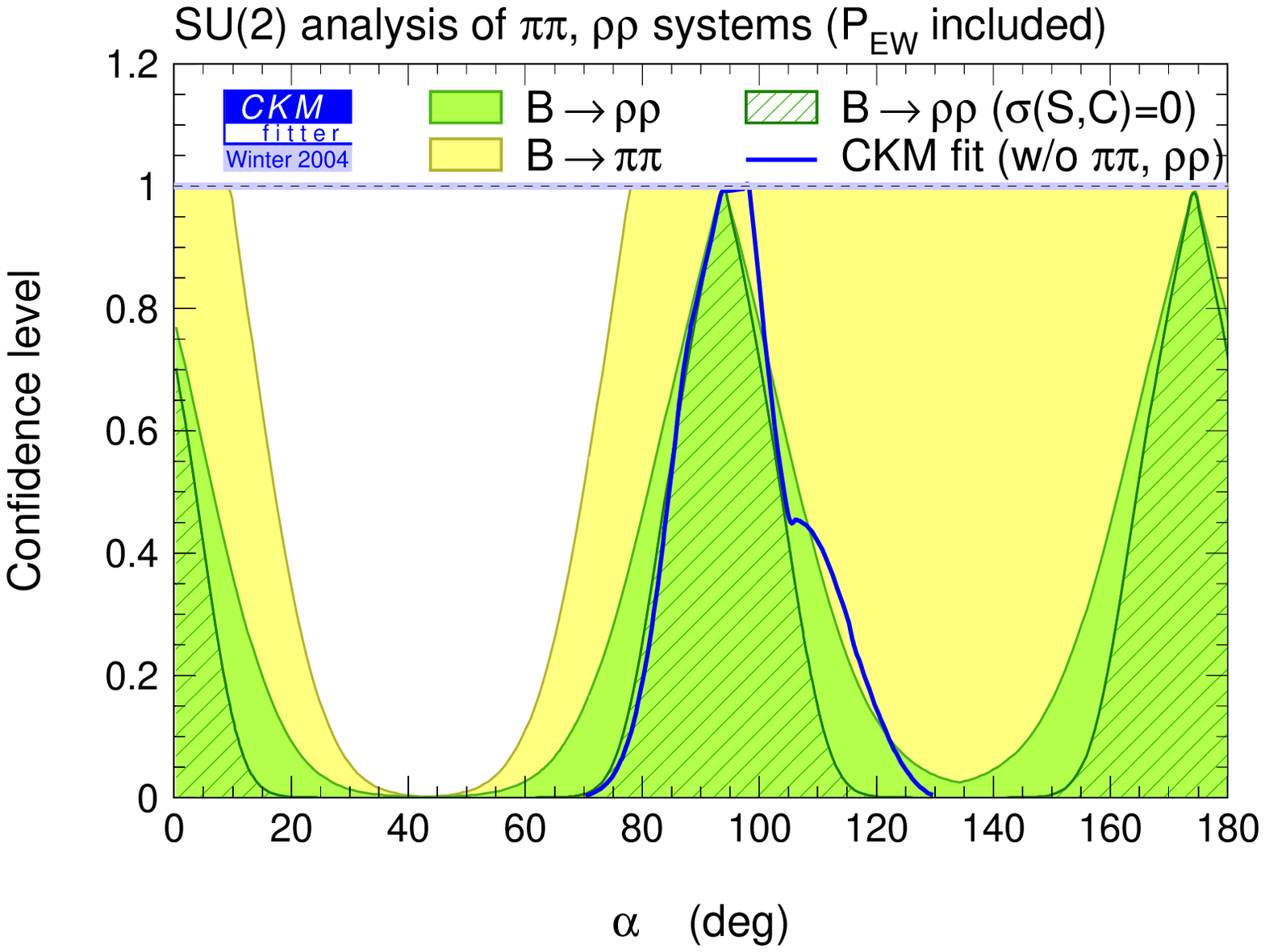}
\end{center}
\end{minipage}\hfill
\begin{minipage}[t]{7cm}
\begin{center}
\includegraphics*[width=7cm]{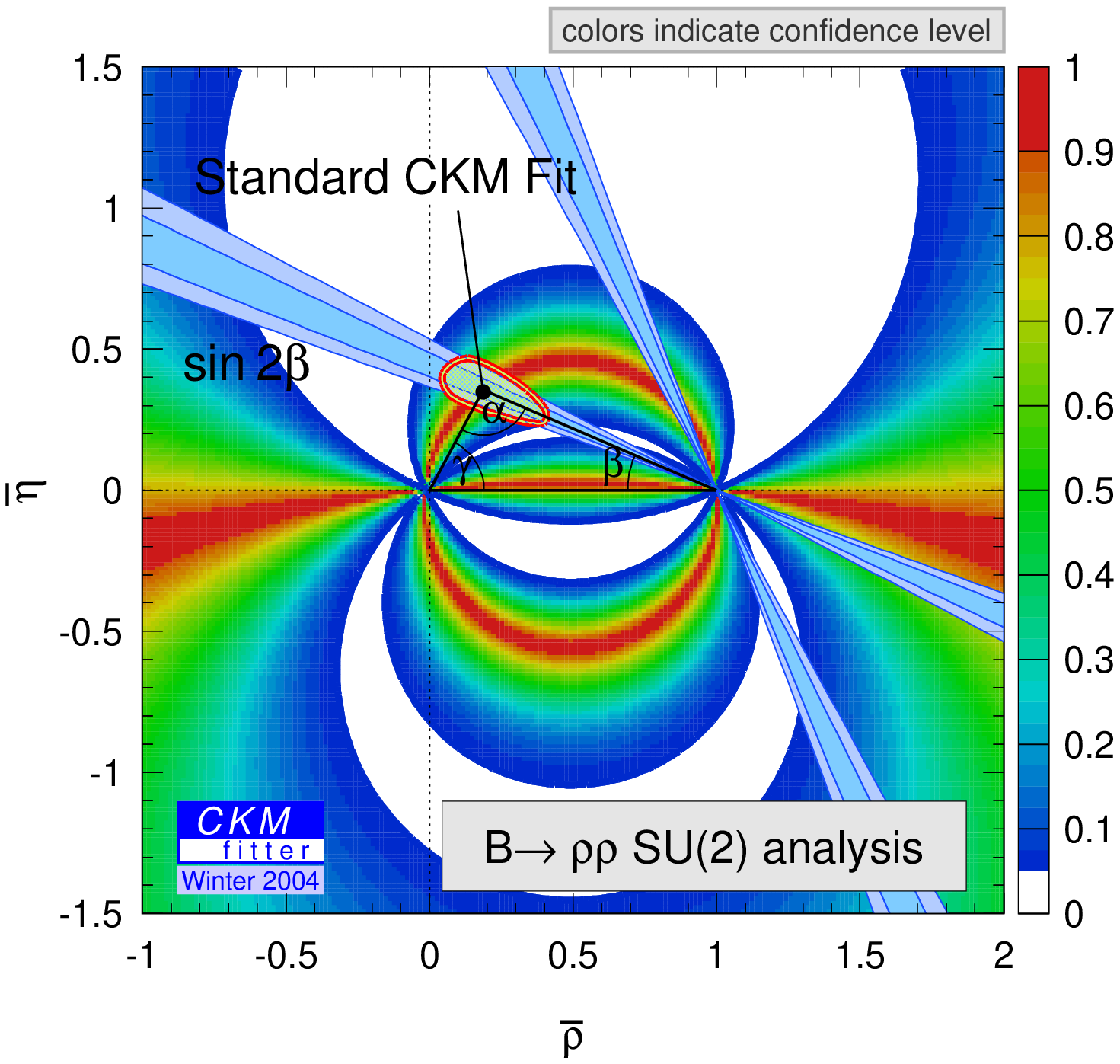}
\end{center}
\end{minipage}
 \caption{
The $B\to\rho\rho$ analysis constrains the possible
values of $\alpha$. The left-hand plot shows the C.L. for different values
of $\alpha$, given the current measurements of $B \to \pi \pi$ from \babar\
and Belle and $B \to \rho \rho$ from \babar\
(with and without experimental errors). 
Overlaid is the global CKM fit without these two analyses 
included.  The right-hand plot shows the constraint on the
$\rho-\eta$ plane due to the $B \to \rho \rho$ analysis,
which is shown overlaid by the Standard CKM fit.}
\label{alpha_pipirr_Wa}
\end{figure}

$B\to \rho^+\rho^-$ is similar to $B\to
\pi^+\pi^-$  but it is a vector--vector decay and can in principle proceed
via three partial waves depending on the angular moments. ``s'' and ``d''
waves have \CP\ even while ``d'' waves have \CP\ odd. Hence, of the three
helicity amplitudes, only the state corresponding to longitudinal
polarization is a pure \CP\ eigenstate. $B\to \rho^+\rho^-$ has been
observed in \babar\ with a BR $(3.0\pm4\pm5)\times10^{-6}$ and with
completely longitudinal polarization, ($99\pm3\pm3$)\%~\cite{rhorho}.
A theoretical limit on the shift between $\alpha$ and $\alpha_{\rm eff}$
is described by  
the Grossman-Quinn bound~\cite{grossman}, which for $B\to\rho\rho$ is written:
\begin{equation}
|\alpha - \alpha_{\rm eff}| < \frac{{\cal B}(B^0\to\rho^0\rho^0)}{{\cal B}(B^0\to\rho^+\rho^-)}.\label{gqbound}
\end{equation}
It provides a
reasonably tight theoretical constraint on the value of 
$|\alpha - \alpha_{\rm eff}|$ of 12.9$\degrees$ at 68.3\% C.L.
From a SU(2) analysis, choosing the result nearest to
the CKM best fit\cite{hagiwara}, we measure $\alpha = (96 \pm 10 \pm 4 \pm
13)\degrees$, where the last error is due to the penguin contamination.
 Fig.~\ref{alpha_pipirr_Wa} shows that
this measurement is more effective than any for the $\pi^+\pi^-$ mode,
that it is
consistent with independent limits from other measurements as found with
CKM fitter\cite{ckmfitter}, and that not much improvement is possible,
even with more statistics, without a better bound on the BR
of $B\to \rho^0\rho^0$.

\subsection{Measurement of $\gamma$}

$\gamma$ measurements can be made in modes which have
both $b\to c$ and $b\to u$ tree diagrams, which interfere.
The magnitude of the interference is determined by the
ratio of the two methods of decay.  

$B^0\to D^{(*)+}\pi^-$ is sensitive to $\sin{(2\beta+\gamma)}$. It is
possible for 
a $B^0$ to decay into $D^{(*)+}\pi^-$ either via a $b\to c$ transition or
via a CKM--suppressed decay with $B$--mixing. The phase $2\beta$ arises
from the mixing and the phase $\gamma$ from the $b\to u$ transition. The
expected asymmetry is small. The analysis is performed with a sample of
fully reconstructed $B$ mesons and a sample of $B$ mesons where the $D^0$
is not explicitly reconstructed. We measure $|\sin{(2\beta+\gamma)}|>
0.58$ (95\% C.L.)~\cite{sin2bpg}. 

\begin{figure}[htb]
\begin{minipage}{0.42\linewidth}
\begin{center}
\includegraphics*[width=0.9\linewidth]{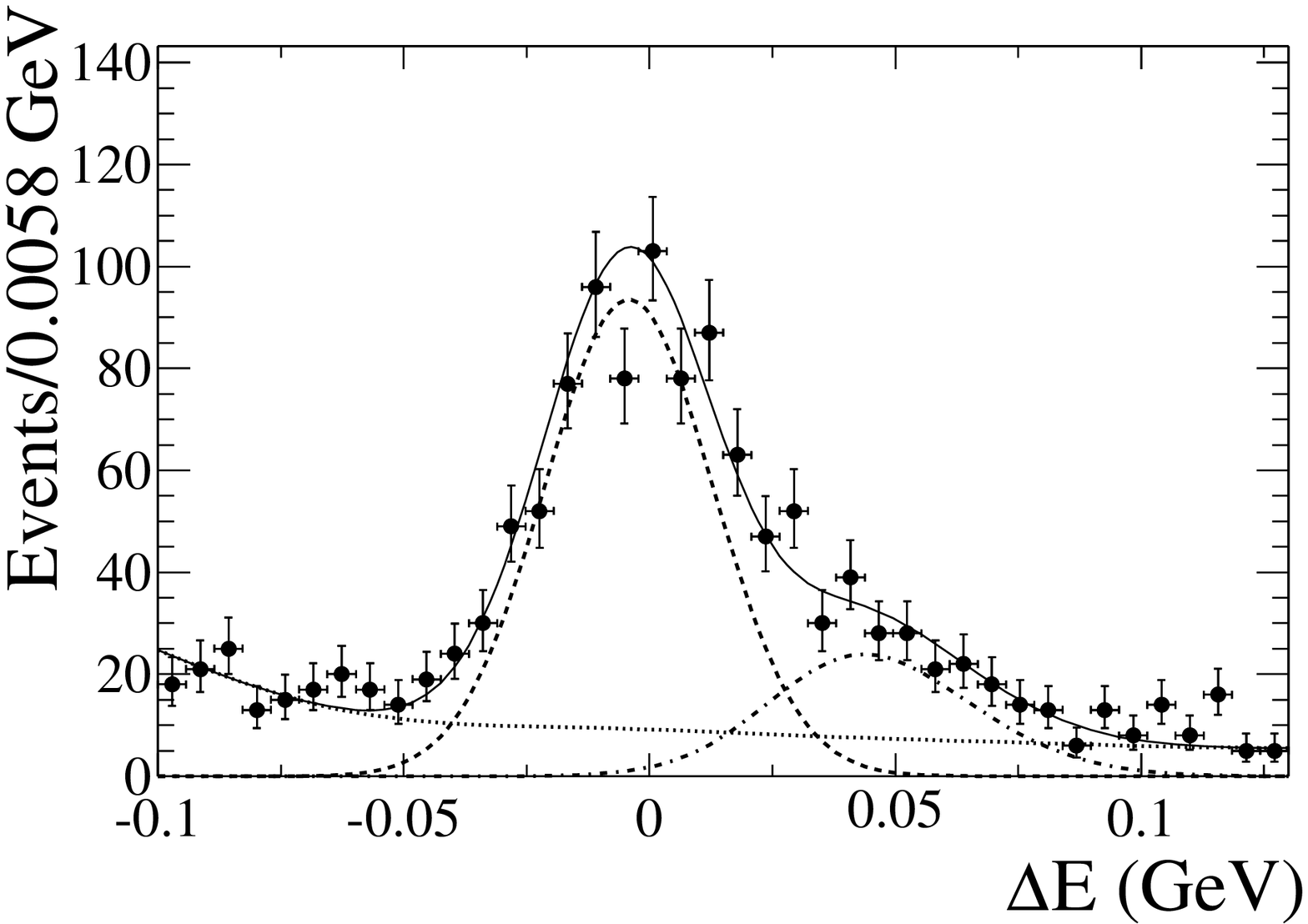}
\end{center}
\end{minipage}\hfill
\begin{minipage}{0.42\linewidth}
\begin{center}
\includegraphics*[width=0.9\linewidth]{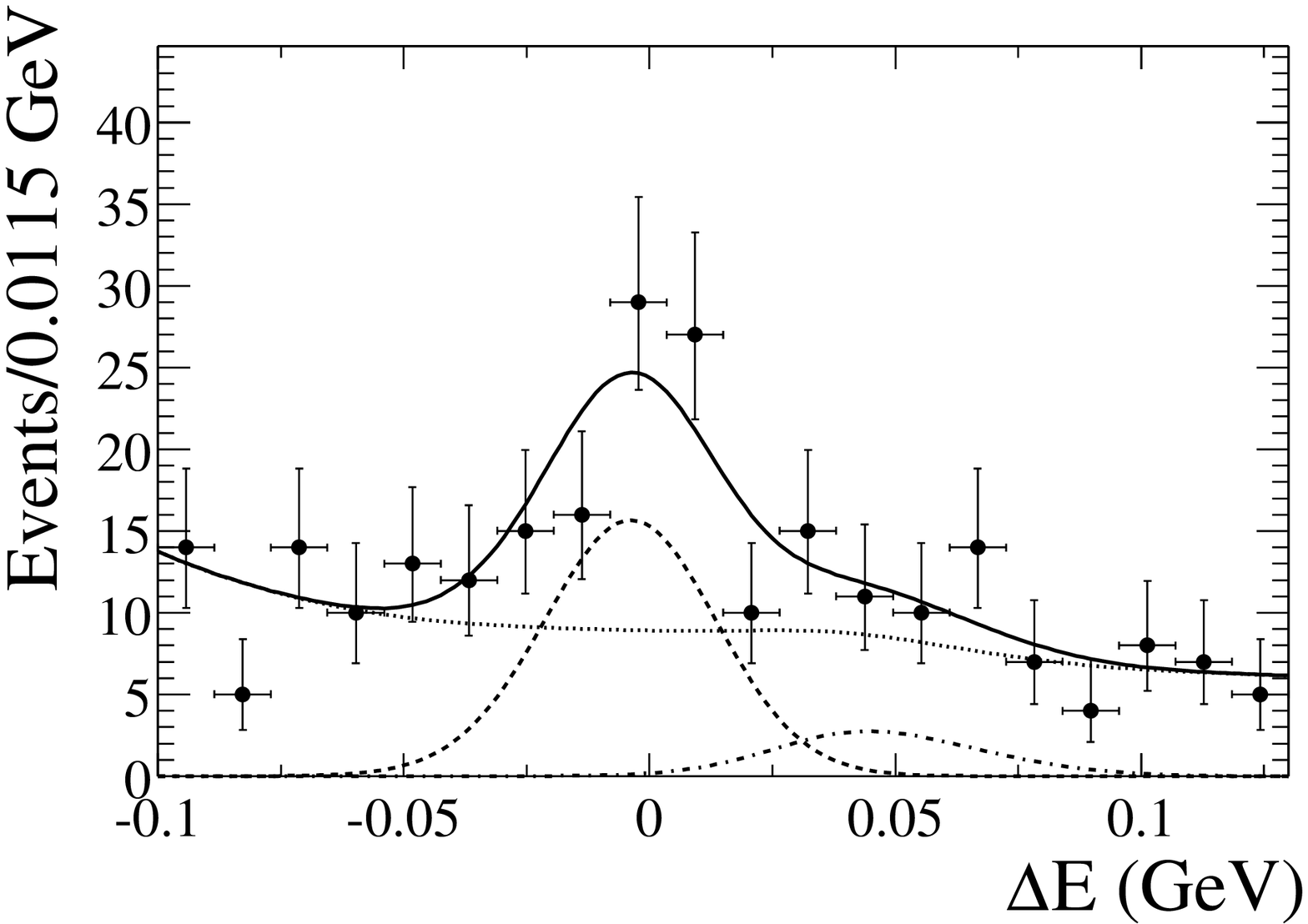}
\end{center}
\end{minipage}\hfill
\caption{\label{fig01a}
$B^-{\ra}D^0K^-$ signal
after requiring that the prompt track be consistent with the
kaon hypothesis for the flavor (left) and \CP\ (right) eigenstates. The
quasi--gaussian peaks on the left (right) of each plot are the
$B^-{\ra}D^0K(\pi)^-$ contributions.} 
\end{figure}

The study of $B^-{\to}D^{(*)0}K^{(*)-}$ decays
will play an important role in our
understanding of \cp\ violation, as they can be used to constrain the
angle $\gamma$  of the 
Cabibbo-Kobayashi-Maskawa (CKM) matrix in a theoretically clean
way~\cite{gronau1991}. 
In the SM, in the absence of $D^{0}\overline{{D}^{0}}$ mixing,
$R_{\CP\pm}/R_{{\rm non}-\CP}\simeq 1+r^2\pm2r\cos\delta \cos\gamma$, where
\begin{equation}
R_{{\rm non}-\CP/\CP\pm}\equiv \frac{\BR(B^-\ra
D^0_{{\rm non}-\CP/\CP\pm}K^-)}{\BR(B^-\ra D^0_{{\rm non}-\CP/\CP\pm}\pi^-)},
\label{eq:rstar}
\end{equation}
$r$ is the ratio of the color suppressed $B^+\ra D^0K^+$ and color
allowed $B^-\ra D^0K^-$ amplitudes ($r \sim
0.1-0.3$), and $\delta$ is the
\cp-conserving strong phase difference between these
amplitudes. Furthermore, defining the direct \CP asymmetry
\begin{equation}
A_{\CP\pm}\equiv \frac{\BR(B^-{\ra}D^0_{\CP\pm}K^-)-\BR(B^+{\ra}D^0_{\CP\pm}K^+)}{\BR(B^-{\ra}D^0_{\CP\pm}K^-)+\BR(B^+{\ra}D^0_{\CP\pm}K^+)},
\label{eq:cpa}
\end{equation}
we have:
$A_{\CP\pm}=\pm 2r\sin\delta\sin\gamma/(1+r^2\pm
2r\cos\delta\cos\gamma)$. The unknowns $\delta$, $r$, and $\gamma$ can be
constrained from the measurements of $R_{{\rm non}-CP}$, $R_{\CP\pm}$, and
$A_{\CP\pm}$. The smaller
$r$ is, the more difficult is the measurement of $\gamma$ with this method. At
\babar\ we have studied the $B^{\pm} \to D^0 K^{\pm}$ 
mode in the flavor ($D^0\to K^-\pi^+, K^-\pi^+\pi^0$, and
$K^-\pi^+\pi^-\pi^+$,  
and the charged conjugate decays) and \CP$=1$ states  ($D^0\to K^+K^-$ and
$\pi^+\pi^-$). 
Two quantities are used to discriminate between signal and background: the
beam-energy-substituted mass 
$\mes \equiv \sqrt{(E_i^{*2}/2 + 
\mathbf{p}_i\cdot\mathbf{p}_B)^2/E_i^2-p_B^2}$
and the energy difference $\Delta E\equiv E^*_B-E_i^*/2$,
where the subscripts $i$ and $B$ refer to the initial 
{\ensuremath{e^+e^-}\xspace}\ system and the  $B$ candidate 
respectively, the asterisk denotes the CM frame, and the kaon mass 
hypothesis of the prompt track is used to calculate $\Delta E$.
Figure~\ref{fig01a} shows the $B^-{\ra}D^0K^-$ signal
after requiring that the prompt track be consistent with the
kaon hypothesis for the flavor and \CP\ eigenstates.
Using datasets of 56\invfb for the measurement of $R$, and 82\invfb for
$R_{CP+}$ and $A_{CP+}$, \babar\ measures\cite{D_CP_K}:
$R = (8.31 \pm 0.35 \pm 0.20)\%$, 
$R_{CP+} = (8.8 \pm 1.6 \pm 0.5)\%$, 
$A_{CP+} =  0.07 \pm 0.17 \pm 0.06$, and  
$R_{CP+}/R  =  1.06 \pm 0.19 \pm 0.06$.  
No meaningful $\gamma$ measurement is yet possible from these results.

We can also use the  Atwood, Dunietz and Soni
method\cite{ads}, which exploits the interference between the decay chain
combining the CKM and color
suppressed $B^+\to D^0K^+$ decay and the CKM allowed $D^0\to K^-\pi^+$
decay and the one with a color allowed $B^+\to \overline{D^0}K^+$ decay
and the doubly CKM suppressed $\overline{D^0}\to K^-\pi^+$ decay. We find no
signal in the suppressed decay mode, and,
using a Bayesian model, we measure: $r < 0.22$ at 90\%
C.L.~\cite{D_Kpi_K}, result which makes a 
measurement of $\gamma$ quite difficult. 

\section{Conclusions}
We measure $\cos 2\beta<0$ at 89\% C.L. and find no evidence of \CPT 
violation, in agreement with SM expectation. Measurements of 
\CP asymmetries in the penguin dominated modes are also found compatible
with SM expectations at the present level of statistics.
The \babar\ experiment has also conducted several analyses with the aim of
extracting $\alpha$ and $\gamma$.  In the $B^0 \to \rho^+ \rho^-$
system, we measure
$\alpha = (96 \pm 10 \pm 4 \pm 13)\degrees$.  Using  $B^0 \to
D^{(*)+}\pi^-$ decays,  
we find $|\sin{(2\beta+\gamma)}| ~>~ 0.58 $
at 95\% C.L. Other decays and methods to extract the angle $\gamma$ are
under investigation, and tighter constraints on its value will be
found once larger data sets become available from both \babar\ and Belle,
though these measurements appear quite hard as \babar\ also
measures: $r<0.22$ at 90\% C.L. 

\bibliographystyle{plain}

\end{document}